\documentclass[pre,aps,preprint,eqsecnum,floatfix]{revtex4}

\usepackage{epsf}
\usepackage{citesort}
\usepackage{bm}
\usepackage{amsmath}
\usepackage{amssymb}
\usepackage{amsgen}
\usepackage{amsfonts}
\usepackage{amsbsy}
\usepackage{version}
\usepackage{url}
\usepackage{wasysym}
\usepackage[dvips]{epsfig,color}

\begin{document}

\newcommand{\rmd}{\mathrm{d}}
\newcommand{\Tr}{\mathrm{Tr}}
\newcommand{\diag}{\mathrm{diag}\,}

\newcommand{\fulltriangleup}{\mbox{$\blacktriangle$}}
\newcommand{\fulltriangledown}{\mbox{$\blacktriangledown$}}
\newcommand{\fullcircle}{\mbox{{\Large$\bullet$}}}
\newcommand{\fullsquare}{\mbox{$\blacksquare$}}
\newcommand{\opentriangle}{\mbox{$\triangle$}}
\newcommand{\opencircle}{\mbox{\Large$\circ$}}
\newcommand{\opensquare}{\mbox{$\Box$}}
\newcommand{\fullline}{\protect\rule[2pt]{15pt}{1pt}}
\newcommand{\dashedline}{\protect\rule[2pt]{3pt}{1pt} \!\protect\rule[2pt]{3pt}{1pt} \!\protect\rule[2pt]{3pt}{1pt}}
 
\renewcommand{\thefigure}{\arabic{section}.\arabic{figure}}\makeatletter

\title{Frustration of nanoconfined liquid crystals due to hybrid substrate anchoring}

\author{Manuel Greschek$^{1}$ and Martin Schoen$^{1,2,}$\footnote{Corresponding author: {\texttt martin.schoen@tu-berlin.de}}}
\affiliation{$^{1}$Stranski-Laboratorium f\"ur Physikalische und Theoretische Chemie,
Fakult\"at f\"ur Mathematik und Naturwissenschaften,
Technische Universit\"at Berlin,
Stra{\ss}e~des~17.~Juni 135, 
10623 Berlin, GERMANY\\
$^{2}$Department of Chemical and Biomolecular Engineering,
Engineering Building I,
Box 7905,
North Carolina State University,
911 Partners Way,
Raleigh, NC 27695, U.S.A.}

\date{\today}

\begin{abstract}
We present Monte Carlo simulations of liquid-crystalline material confined to a nanoscopic slit-pore. The simulations are carried out under isothermal conditions in a specialized isostress ensemble in which $N$ fluid molecules are exposed to a compressional stress $\tau_{\parallel}$ acting on the fluid in directions parallel with the substrate planes. Fluid-fluid and fluid-substrate interactions are modelled as in our previous work [M. Greschek {\em et al.}, {\em Soft Matter}, 2010, DOI:10.1039/B924417D). We study several anchoring mechanisms at the solid substrate by introducing an anchoring function $0\le g\left(\widehat{\bm{u}}\right)\le1$ that depends on the orientation $\widehat{\bm{u}}$ of a fluid molecule relative to the substrate plane; $g\left(\widehat{\bm{u}}\right)$ ``switches'' the fluid-substrate {\em attraction} on or off. Here we focus on various {\em heterogeneous} (i.e., hybrid) anchoring scenarios imposing different anchoring functions at the opposite substrates. As in our previous study we compute the isostress heat capacity which allows us to identify states at which the confined fluid undergoes a structural transformation. The isotropic-nematic transformation turns out to be nearly independent of the specific anchoring scenario. This is because the director in the nematic phase assumes a direction that is a compromise between the directions enforced by the competing anchoring scenarios at either substrate. On the contrary, at lower compressional stresses molecules prealign in specific directions that depend on details of the anchoring scenario. 
\end{abstract}

\maketitle

\section{Introduction}\label{sec:intro}
If a liquid crystal is in its nematic phase specific orientations of the director [i.e., the (unit) vector $\widehat{\bm{n}}$ with which molecules align preferentially] can be realized through specific anchoring scenarios at the solid substrate. The substrate-induced alignment can be perturbed if an external field is superimposed such that it is competing with the director. An example is the so-called Fr\'eedericksz cell in which a nematic liquid crystal is exposed to a magnetic field pointing in a direction other than that in which molecules are anchored at the walls of the cells \cite{freedericksz33,ziherl00}. If the external field is sufficiently strong and if the substrate separation is large enough there will be a region in the liquid crystal where the local director points in another direction than that enforced by substrate anchoring. In other words, the orientational order in the nematic phase is perturbed by the presence of the external field. Because the correlation lengths of these perturbations may be substantial they can be used in sensor applications. This was recently demonstrated by Guzm\'an {\em et al.} who considered a model for a bionanosensor based upon a confined liquid-crystalline phase \cite{guzman05}. The sensor consists of thin films of liquid-crystalline materials confined between substrate surfaces with specific anchoring characteristics. If particles such as proteins or viruses bind to these solid surfaces they give rise to long-range perturbations of the orientational order in a nematic phase that reports the binding of these particles at specific receptors on the solid surfaces \cite{gupta98,skaife00,skaife01}. The binding process perturbs the local structure of the liquid-crystal phase over sufficiently long distances so that optical methods can be used to detect the perturbation.

The apparent importance of various anchoring scenarios for the nematic phase of such nanoconfined liquid crystals prompted us to investigate the impact of various anchoring scenarios on the isotropic-nematic (IN) phase transition in a model liquid crystal of prolate ellipsoids of revolution confined to a nanoscopic slit-pore \cite{greschek10}. Our results indicate that the nature of the anchoring at the substrate has a pronounced effect on the location of the IN transition and on structural details associated with that transition. In view of our earlier results it thus seems interesting to replace in this work the original {\em homogeneous} \cite{greschek10} by {\em heterogeneous} (or ``hybrid'') anchoring scenarios such that the anchoring is different at the opposite substrates forming the slit-pore. This will cause the director field to be inhomogeneous such that interesting new physical phenomena are anticipated.

That such heterogeneous or competing fluid-wall interactions can lead to exciting effects and interesting new physics has already been demonstrated in the past in several contexts. Perhaps the best-known example is the localization-delocalization transition that occurs for the gas-liquid interface in fluids confined between two planar walls where one wall favors the liquid- and the other one favors the gas-like phase \cite{binder95,binder96,ferrenberg98}. If the substrate is nonplanar such as in the bi-pyramid considered by Milchev {\em et al.} \cite{milchev05} competing surface fields can cause the spontaneous magnetization in an Ising magnet to vanish at a filling temperature $T_{\mathrm{f}}$ below the bulk critical temperature $T_{\mathrm{cb}}$; the spontaneous magnetization remains zero for all $T\ge T_{\mathrm{f}}$.

In the context of liquid crystals surprisingly little attention has been paid so far to hybrid anchoring scenarios despite their importance for optical applications \cite{memmer03,chrzanowska01}. For example, Ziherl {\em et al.} consider a situation in which the molecules of a liquid crystal are aligned homeotropically at one wall and planar at the opposite such that in the nematic phase molecules appear to be frustrated with respect to the direction of $\widehat{\bm{n}}$ \cite{ziherl00}. The authors focus on the impact of this frustration on fluctuation-induced forces that can be interpreted as an analogue of the well-known Casimir forces in quantum systems \cite{krech94}. For the same hybrid anchoring scenario Rodr\'iguez-Ponce {\em et al.} employ mean-field density functional theory and observe a linear variation of the tilt angle between the local director field and the substrate normal (see also Ref.~\citealp{priezjew03}) as far as strong anchoring at the substrate is concerned \cite{rodriguez01}. Such a linear variation of the tilt-angle profile has also been reported by Steuer {\em et al.} \cite{steuer04} for hybrid anchoring corresponding to a nanoscopic twisted nematic cell and for the same model system used in this work. However, to date a systematic study of the impact of hybrid anchoring on the IN transition is still lacking.

Experimentally, hybrid anchoring scenarios can be realized in a number of ways. For example, Chung {\em et al.} use photoalignment to fabricate planar substrates where the anchoring at each substrate differs by an angle of $\pi/4$ \cite{chung02}. They show that the resulting molecular orientation in a confined liquid crystal can be controlled by the photoirradiation time. The same authors show that very similar results can be obtained if the surfaces are rubbed in specific directions and under controlled conditions. Zappone {\em et al.} perform experiments using the surface forces apparatus (SFA) to measure the solvation force in nematic liquid crystals under hybrid anchoring conditions \cite{zappone05}. They set up a hybrid (homeotropic-planar) anchoring scenario by dipping the muscovite mica surfaces, with which the crossed cylinders in the SFA setup are coated, in various cationic surfactant solutions where the surfactants differ only in the length of the aliphatic chains.

In this work we will study the impact of various hybrid anchoring scenarios on properties of nanoconfined model liquid crystals by means of Monte Carlo simulations in a specialized isostress ensemble similar to our previous study \cite{greschek10}. Our emphasis will be on structural changes arising in the confined phase as the transverse compressional stress $\tau_{\parallel}$ increases. Increasing $\tau_{\parallel}$ eventually drives the system from an isotropic to a nematic phase. We have organized the remainder of our manuscript such that we will briefly describe the model system in Sec.~\ref{sec:mod}. Our results are presented in Sec.~\ref{sec:res} and summarized and discussed in the concluding Sec.~\ref{sec:sumcon}. Because of the hybrid anchoring scenarios employed in this study biaxiality of the nematic phases becomes an issue. The biaxial order parameter is introduced in Appendix~\ref{sec:appa} using perturbation theory. 
\setcounter{figure}{0}
\section{The model system}\label{sec:mod}
\subsection{Interaction potentials}\label{sec:intpot}
The model liquid crystal considered in this work is confined to a mesoscopic slit-pore such that we may decompose the total configurational energy into a fluid-fluid (ff) and into a fluid-substrate (fs) contribution according to
\begin{equation}\label{eq:u}
U(\bm{R},\widehat{\bm{U}})=
U_{\mathrm{ff}}(\bm{R},\widehat{\bm{U}})+
U_{\mathrm{fs}}(\bm{R},\widehat{\bm{U}})
\end{equation}
where $\bm{R}\equiv\left\{\bm{r}_1,\bm{r}_2,\ldots,\bm{r}_N\right\}$ and $\widehat{\bm{U}}\equiv\left\{\widehat{\bm{u}}_1,\widehat{\bm{u}}_2,\ldots,\widehat{\bm{u}}_n\right\}$ are shorthand notations for the sets of center-of-mass coordinates and unit vectors specifying the orientations of the $N$ liquid-crystalline molecules, respectively. More specifically,
\begin{equation}\label{eq:uff}
U_{\mathrm{ff}}(\bm{R},\widehat{\bm{U}})=
\frac{1}{2}
\sum\limits_{i=1}^{N}
\sum\limits_{j\ne i}^{N}
u_{\mathrm{ff}}\left(\bm{r}_{ij},\widehat{\bm{u}}_i,\widehat{\bm{u}}_j\right)
\end{equation}
where $\bm{r}_{ij}\equiv\bm{r}_i-\bm{r}_j$ is the distance vector between the centers of mass of particles $i$ and $j$ assuming pairwise additivity of their interactions. As in our previous work \cite{greschek10} we take $u_{\mathrm{ff}}$ as (see also Ref.~\citealp{hess99})
\begin{equation}\label{eq:ljorient}
u_{\mathrm{ff}}\left(\bm{r}_{ij},\widehat{\bm{u}}_i,\widehat{\bm{u}}_j\right)=
4\varepsilon
\left[
\left(
\frac{\sigma}{r_{ij}}
\right)^{12}-
\left(
\frac{\sigma}{r_{ij}}
\right)^{6}
\left\{
1+\Psi\left(\widehat{\bm{r}}_{ij},\widehat{\bm{u}}_i,\widehat{\bm{u}}_j\right)
\right\}
\right]
\end{equation}
where $r=\left|\bm{r}\right|$, and $\widehat{\bm{r}}=\bm{r}/r$. Hence, $u_{\mathrm{ff}}$ is a Lennard-Jones potential where the attractive contribution is modified to account for the orientation dependence of the interaction between a pair of molecules. In eqn.~(\ref{eq:ljorient}), $\sigma$ denotes the ``diameter'' of a spherical reference molecule and $\varepsilon$ is the depth of the attractive well in that reference model. The anisotropy of the fluid-fluid interaction is accounted for by the function
\begin{equation}\label{eq:psi}
\Psi\left(\widehat{\bm{r}}_{ij},\widehat{\bm{u}}_i,\widehat{\bm{u}}_j\right)=
5\varepsilon_1
P_2\left(\widehat{\bm{u}}_i\cdot\widehat{\bm{u}}_j\right)+
5\varepsilon_2
\left[
P_2\left(\widehat{\bm{u}}_i\cdot\widehat{\bm{r}}_{ij}\right)+
P_2\left(\widehat{\bm{u}}_j\cdot\widehat{\bm{r}}_{ij}\right)
\right]
\end{equation}
where we take $\varepsilon_1=0.04$ and $\varepsilon_2=-0.08$ as in our previous study \cite{greschek10}. As we have demonstrated there (see Fig.~1 of Ref.~\citealp{greschek10}) our molecules turn out to be prolate ellipsoids of revolution with an aspect ratio corresponding to $1.26$. The specific functional form of $\Psi$ preserves the head-tail symmetry of our molecules, that is orientations $\widehat{\bm{u}}_i$ and $-\widehat{\bm{u}}_i$ are equivalent. In eqn.~(\ref{eq:psi})
\begin{equation}
P_2\left(x\right)=
\frac{1}{2}
\left(
3x^2-1
\right)
\end{equation}
is the second Legendre polynomial.

To model the fluid-substrate contribution to $U$ we follow earlier work \cite{steuer04,greschek10} and introduce
\begin{equation}
U_{\mathrm{fs}}(\bm{R},\widehat{\bm{U}})=
\sum\limits_{k=1}^2
\sum\limits_{i=1}^N
u_{\mathrm{fs}}^{\left[k\right]}\left(\bm{r}_i,\widehat{\bm{u}}_i\right)
\end{equation}
where
\begin{equation}\label{eq:smoothwall}
u_{\mathrm{fs}}^{\left[k\right]}\left(z_i,\widehat{\bm{u}}_i\right)=
2\pi\varepsilon\rho_{\mathrm{s}}\sigma^2
\left[
\frac{2}{5}
\left(
\frac{\sigma}{z_i\pm s_{\mathrm{z}0}/2}
\right)^{10}-
\left(
\frac{\sigma}{z_i\pm s_{\mathrm{z}0}/2}
\right)^{4}
g^{\left[k\right]}\left(\widehat{\bm{u}}_i\right)
\right]
\end{equation}
In eqn.~(\ref{eq:smoothwall}), $z_i$ is the $z$-coordinate of the center-of-mass position of molecule $i$ in a space-fixed Cartesian coordinate system. In this coordinate system we assume the lower substrate ($k=1$) to be located at $z_{\mathrm{w}}=-s_{\mathrm{z}0}/2$ whereas the upper one is located at  $z_{\mathrm{w}}=+s_{\mathrm{z}0}/2$ ($k=2$). In other words, eqn.~(\ref{eq:smoothwall}) assumes the substrates to be structureless such that at fixed molecular orientation $u_{\mathrm{fs}}$ depends only on the distance of molecule $i$ from either substrate along the $z$-axis. In the prefactor the areal density of the solid substrates $\rho_{\mathrm{s}}=2/\ell^2$ where $\ell/\sigma=\sqrt[3]{4}$ is the lattice constant of a single ($100$) plane of a  face-centred cubic lattice.

\subsection{Anchoring scenarios}\label{sec:anch}
In eqn.~(\ref{eq:smoothwall}), $0\le g^{\left[k\right]}\left(\widehat{\bm{u}}_i\right)\le1$ is the so-called anchoring function. Choosing different functional forms the anchoring function permits to realize different, energetically favorable orientations of molecule $i$ with respect to the substrate plane. Specifically, we take $g^{\left[k\right]}\left(\widehat{\bm{u}}_i\right)$ to be given by one of the following expressions:
\begin{subequations}\label{eq:anch}
\begin{eqnarray}
g_0\left(\widehat{\bm{u}}\right)&=&1\label{eq:nonspecific}\\
g_{\perp}\left(\widehat{\bm{u}}\right)&=&\left(\widehat{\bm{u}}\cdot\widehat{\bm{e}}_{\mathrm{z}}\right)^2\label{eq:homeotropic}\\
g_{\parallel}\left(\widehat{\bm{u}}\right)&=&
\left(\widehat{\bm{u}}\cdot\widehat{\bm{e}}_{\mathrm{x}}\right)^2+
\left(\widehat{\bm{u}}\cdot\widehat{\bm{e}}_{\mathrm{y}}\right)^2\label{eq:parallel}\\
g_{\mathrm{x}}\left(\widehat{\bm{u}}\right)&=&\left(\widehat{\bm{u}}\cdot\widehat{\bm{e}}_{\mathrm{x}}\right)^2\label{eq:directional}
\end{eqnarray}
\end{subequations}

The anchoring functions given in eqn.~(\ref{eq:anch}) have the following physical significance. As we rationalize elsewhere \cite{greschek10}, if $g^{\left[k\right]}\left(\widehat{\bm{u}}\right)$ is given by $g_0\left(\widehat{\bm{u}}\right)$ molecules orient themselves in a homeotropic fashion at the substrates even though the substrates themselves do not discriminate any molecular orientation energetically. Employing the expression given in eqn.~(\ref{eq:homeotropic}) also results in homeotropic alignment which is, however, stronger than the one induced by $g_0\left(\widehat{\bm{u}}\right)$ because the substrates discriminate the homeotropic alignment directly. To overcome the homeotropic alignment already supported by the mere presence of the substrate we also consider the anchoring functions specified in eqn.~(\ref{eq:parallel}) and (\ref{eq:directional}). In both cases a planar orientation parallel with the substrate plane is energetically favored. The difference is that by employing eqn.~(\ref{eq:parallel}) {\em any} orientation of a molecule parallel with the substrate plane is energetically favorable whereas eqn.~(\ref{eq:directional}) selects those orientations where $\widehat{\bm{u}}$ is aligned with the $x$-axis. We shall distinguish {\em homogeneous} anchoring scenarios where $g^{\left[1\right]}\left(\widehat{\bm{u}}_i\right)=g^{\left[2\right]}\left(\widehat{\bm{u}}_i\right)$ from {\em heterogeneous} (i.e., ``hybrid'') ones characterized by $g^{\left[1\right]}\left(\widehat{\bm{u}}_i\right)\ne g^{\left[2\right]}\left(\widehat{\bm{u}}_i\right)$ (see Table~\ref{tab1}).

\begin{table}[htb]
\caption{Hybrid anchoring scenarios employed in this work.}
\label{tab1}
\begin{center}
\begin{tabular}{c|c|c|c}
\hline\hline
\multicolumn{2}{c}{$g^{\left[k\right]}\left(\widehat{\bm{u}}\right)$} \vline& \multicolumn{2}{c}{}\\
\hline
$k=1$ & $k=2$ & anchoring scenario& acronym\\
\hline
$g_{\perp}\left(\widehat{\bm{u}}\right)$ & $g_{\parallel}\left(\widehat{\bm{u}}\right)$ & homeotropic-planar & {\em hp}\\
$g_0\left(\widehat{\bm{u}}\right)$ & $g_{\parallel}\left(\widehat{\bm{u}}\right)$ & nonspecific-planar & {\em np}\\
$g_0\left(\widehat{\bm{u}}\right)$ & $g_{\mathrm{x}}\left(\widehat{\bm{u}}\right)$ & nonspecific-directional & {\em nd}\\
\hline\hline
\end{tabular}
\end{center}
\end{table}
\setcounter{figure}{0}
\section{Results}\label{sec:res}
\subsection{Numerical details}\label{sec:numdet}
As in our previous study \cite{greschek10} we employ Monte Carlo simulations in a specialized isostress ensemble in which the fluid is exposed to a constant compressional stress in directions parallel with the substrate planes. We employ periodic boundary conditions in the $x$- and $y$-directions. To generate a Markov chain of configurations in this ensemble we utilize the algorithm described in Sec.~4.1 of Ref.~\citealp{schoen99}. We refer to a Monte Carlo cycle as a sequence of $N$ attempted displacements and rotations of sequentially selected fluid particles plus one attempted change of the area $A$ of the computational cell in the $x$--$y$ plane. Typical runs comprise $10^{5}$ Monte Carlo cycles in regions where the fluid does not undergo an isotropic-nematic (IN) transformation. In the immediate vicinity of a structural transformation the length of a typical run was enlarged to $3\times10^{6}$ Monte Carlo cycles to guarantee sufficient statistical accuracy of the results (see Sec.~IV~B. of Ref.~\citealp{greschek10}).

\begin{figure}[htb]
\epsfig{file=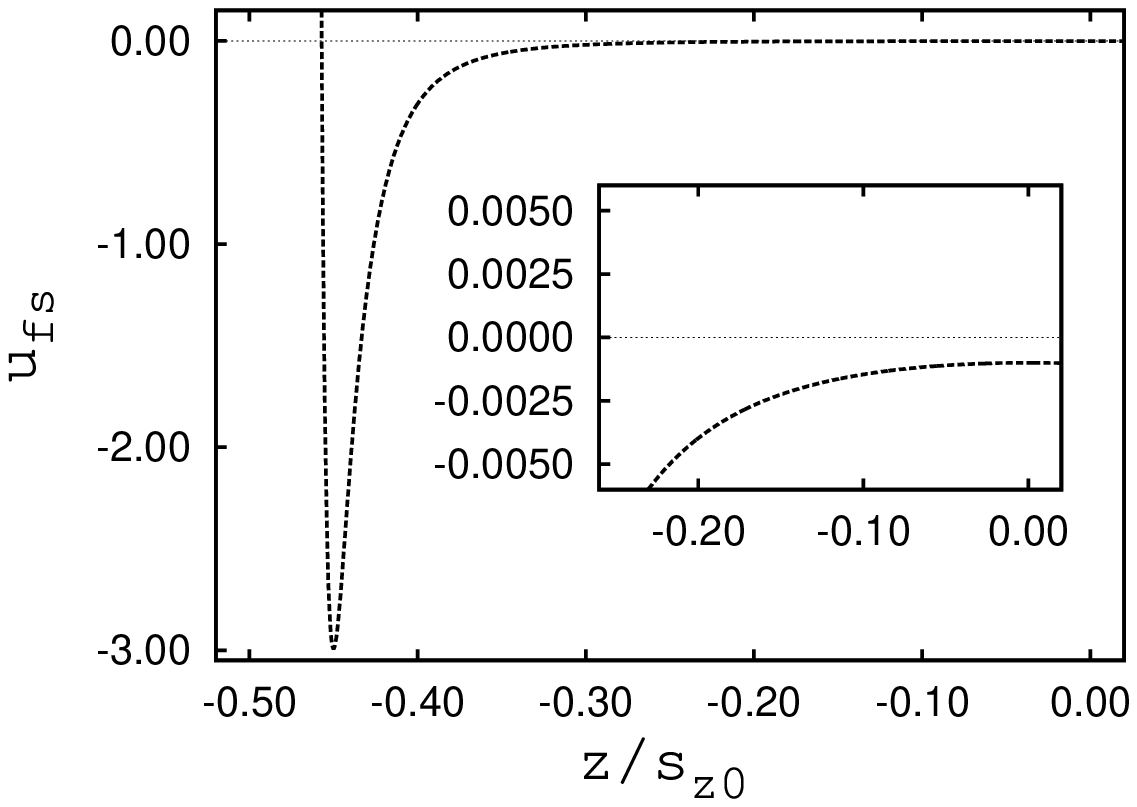,width=0.6\linewidth}
\caption{\small Plot of $u_{\mathrm{fs}}\left(z,\widehat{\bm{u}}\right)=u_{\mathrm{fs}}^{\left[1\right]}\left(z,\widehat{\bm{u}}\right)+u_{\mathrm{fs}}^{\left[2\right]}\left(z,\widehat{\bm{u}}\right)$ as a function of position $z/s_{\mathrm{z}0}$ in the lower half of the system ($z/s_{\mathrm{z}0}\le-0.5$) for $s_{\mathrm{z}0}=20$. The plots are symmetric with respect to the line $z=0$ and were obtained assuming $g^{\left[1\right]}\left(\widehat{\bm{u}}\right)=g^{\left[2\right]}\left(\widehat{\bm{u}}\right)=1$ The inset shows an enlargement of the plot.}\label{fig1}
\end{figure}

We express all quantities of interest in terms of the customary dimensionless (i.e., ``reduced'') units. For example, length is given in units of $\sigma$, energy in units of $\varepsilon$, and temperature in units of $\varepsilon/k_{\mathrm{B}}$. Other derived quantities are expressed in terms of suitable combinations of these basic quantities. For example, stress is given in units of $\varepsilon/\sigma^{3}$. Throughout this work we fix the temperature $T=1.0$ which should be sufficiently subcritical taking the mean-field phase diagram of Hess and Su as a rough guidance \cite{hess99}. Our systems comprise $N=3000$ fluid molecules which we equilibrate during an intial  $10^4$ Monte Carlo cycles. To save computer time we cut off fluid-fluid interactions beyond a separation of $r_{\mathrm{c}}=3.0$ between the centers-of-mass of a pair of fluid molecules. In addition, we employ a linked-cell in combination with a conventional (Verlet) neighbor list as described in the book by Allen and Tildesley to further speed up the simulations (see Chap.~5.3 of Ref.~\citealp{allen86}). This list includes all particles as neighbors whose centers-of-mass are located within a distance of $r_{\mathrm{n}}=3.5$ from a reference molecule at the origin of the neighbor sphere.

The simulations of this work are carried out for a substrate separation $s_{\mathrm{z}0}=20$. This value has been chosen because then our confined fluid contains a sufficiently large region centered around the middle of the slit-pore in which fluid molecules do not interact with either of the two substrates. This is illustrated by plots in Fig.~\ref{fig1} indicating that $u_{\mathrm{fs}}^{\left[1\right]}\left(z,\widehat{\bm{u}}\right)+u_{\mathrm{fs}}^{\left[2\right]}\left(z,\widehat{\bm{u}}\right)\lesssim3\cdot10^{-3}$ for $\left|z/s_{\mathrm{z}0}\right|\lesssim0.20$.

\subsection{Properties}\label{sec:prop}
To investigate the impact of the various hybrid anchoring scenarios introduced in Sec.~\ref{sec:anch} we introduce in the following some key properties on which we shall base our subsequent discussion. The most important one is the so-called alignment tensor \cite{pardowitz80}
\begin{equation}\label{eq:alignmenttensor}
\mathbf{Q}\equiv
\frac{1}{2N}
\sum\limits_{i=1}^N
\left(
3\widehat{\bm{u}}_i\otimes\widehat{\bm{u}}_i-\mathbf{1}
\right)
\end{equation}
where ``$\otimes$'' denotes the direct (i.e., dyadic) product and $\mathbf{1}$ is the unit tensor. The alignment tensor can be represented by a real, symmetric, traceless, $3\times3$ matrix that can be diagonalized numerically using Jacobi's method \cite{press89}. For a system with biaxial symmetry it can be shown that the diagonalized alignment tensor can be written as  [see eqn.~(\ref{eq:diagQ})]  
\begin{equation}\label{eq:biax}
\diag\mathbf{Q}\equiv
\bm{\Lambda}=
\left(
\begin{array}{ccc}
-\lambda_{\mathrm{m}}/2-\zeta&0&0\\
0&-\lambda_{\mathrm{m}}/2+\zeta&0\\
0&0&\lambda_{\mathrm{m}}
\end{array}
\right)
\end{equation}
This expression agrees with the one derived by Low who uses an expansion in terms of Wigner matrices \cite{low02}. Following previous workers \cite{eppenga84,richter06,greschek10} we adopt $S\equiv\left\langle\lambda_{\mathrm{m}}\right\rangle$ as a definition of the Maier-Saupe nematic order parameter \cite{maier59,maier60} where the angular brackets denote an ensemble average in the specialized isostress ensemble. Once the largest eigenvalue of $\mathbf{Q}$ is known one may compute $\zeta$ from eqn.~(\ref{eq:biax}) and with it the biaxial order parameter $\xi\equiv\left\langle\zeta\right\rangle$.

Other structural quantities that we shall be considering below take notice of the anisotropy and inhomogeneity of liquid crystals in confinement. The simplest one of these is the local density of the confined fluid defined as
\begin{equation}\label{eq:rhoz}
\rho\left(z\right)=
\left\langle
\sum\limits_{i=1}^N
\delta\left(z-z_i\right)
\right\rangle=
\frac{1}{\delta z}
\left\langle
\frac{N\left(z\right)}{A}
\right\rangle
\end{equation}
where $\delta\left(z-z_i\right)$ is the Dirac $ \delta$-function and $N(z)$ is the number of molecules within a small interval $\delta z$ centered on $z$. In addition, we consider the {\em local} nematic order parameter $S(z)$ which we compute from an expression for $\mathbf{Q}\left(z\right)$ analogous to eqn.~(\ref{eq:alignmenttensor}) replacing, however, $\widehat{\bm{u}}_i$ and $N$ by their local counterparts $\widehat{\bm{u}}_i(z)$ and $N(z)$, respectively. However, to gain even deeper insight into the nature of nematic phases it turns out to be useful to consider another measure of local nematic order provided by the (local value of the) second Legendre polynomial
\begin{equation}\label{eq:pz}
P_{\mathrm{z}}\left(z\right)\equiv
\frac{1}{2}
\left\langle
\frac{1}{N\left(z\right)}
\sum\limits_{i=1}^{N\left(z\right)}
\left[
3\left(\widehat{\bm{u}}_i\cdot\widehat{\bm{e}}_{\mathrm{z}}\right)^2-1
\right]
\right\rangle
\end{equation}
Hence, $P_{\mathrm{z}}\left(z\right)=1$ if all molecules are homeotropically aligned ($\left|\widehat{\bm{u}}_i\cdot\widehat{\bm{e}}_{\mathrm{z}}\right|=1$). If, on the other hand, molecules are aligned in a parallel fashion $P_{\mathrm{z}}\left(z\right)=-0.5$ ($\left|\widehat{\bm{u}}_i\cdot\widehat{\bm{e}}_{\mathrm{z}}\right|=0$). Notice, that in writing eqn.~(\ref{eq:pz}) we {\em assume} $\widehat{\bm{n}}$ to point along the $z$-axis of a space-fixed Cartesian coordinate system whereas the true direction of the {\em local} director $\widehat{\bm{n}}\left(z\right)$ computed as an eigenvector of $\mathbf{Q}\left(z\right)$ is unrestricted.

In addition to these structural quantities we compute a specialized isostress heat capacity
\begin{equation}\label{eq:ctaufinal}
c_{\tau}=
\frac{5}{2}k_{\mathrm{B}}+
\frac{\left\langle\mathcal{H}^2\right\rangle-
\left\langle\mathcal{H}\right\rangle^2}{Nk_{\mathrm{B}}T^2}
\end{equation}
that we already introduced in our previous study \cite{greschek10}. As we have demonstrated there, $c_{\tau}$ is particularly useful to identify characteristic stresses at which the confined liquid crystal may undergo subtle structural transformations. In eqn.~(\ref{eq:ctaufinal})
\begin{equation}\label{eq:enthalpy}
\mathcal{H}(\bm{R},\widehat{\bm{U}})\equiv
U(\bm{R},\widehat{\bm{U}})-
\tau_{\parallel}As_{\mathrm{z}0}
\end{equation}
such that $\left\langle\mathcal{H}\right\rangle$ is a specialized enthalpy.

\subsection{Heat capacity and global order parameters}\label{sec:hc}
\begin{figure}
\begin{center}
\epsfig{file=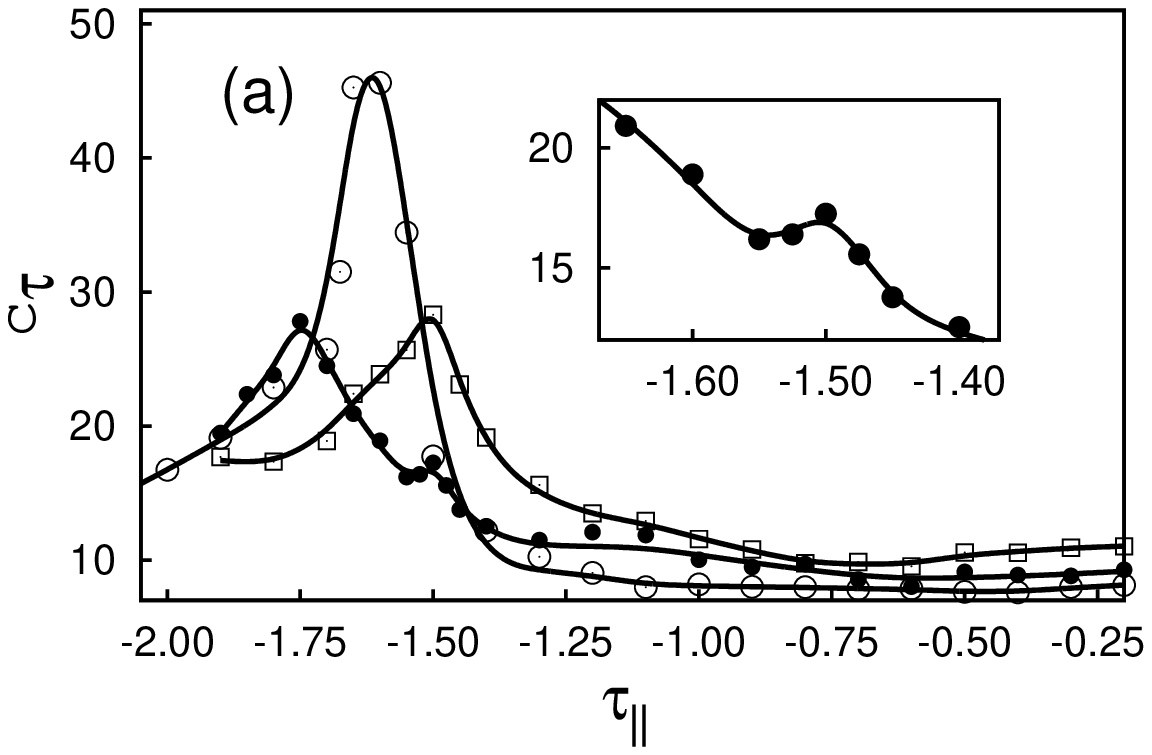,width=0.55\linewidth}
\epsfig{file=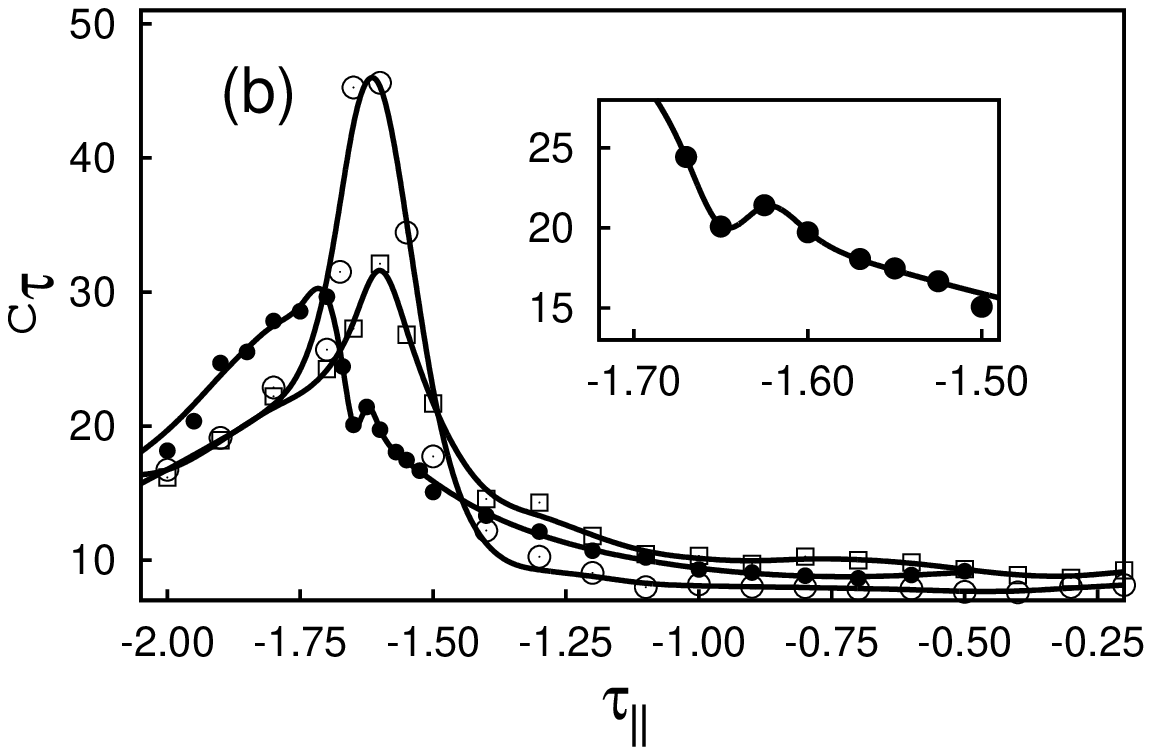,width=0.55\linewidth}
\epsfig{file=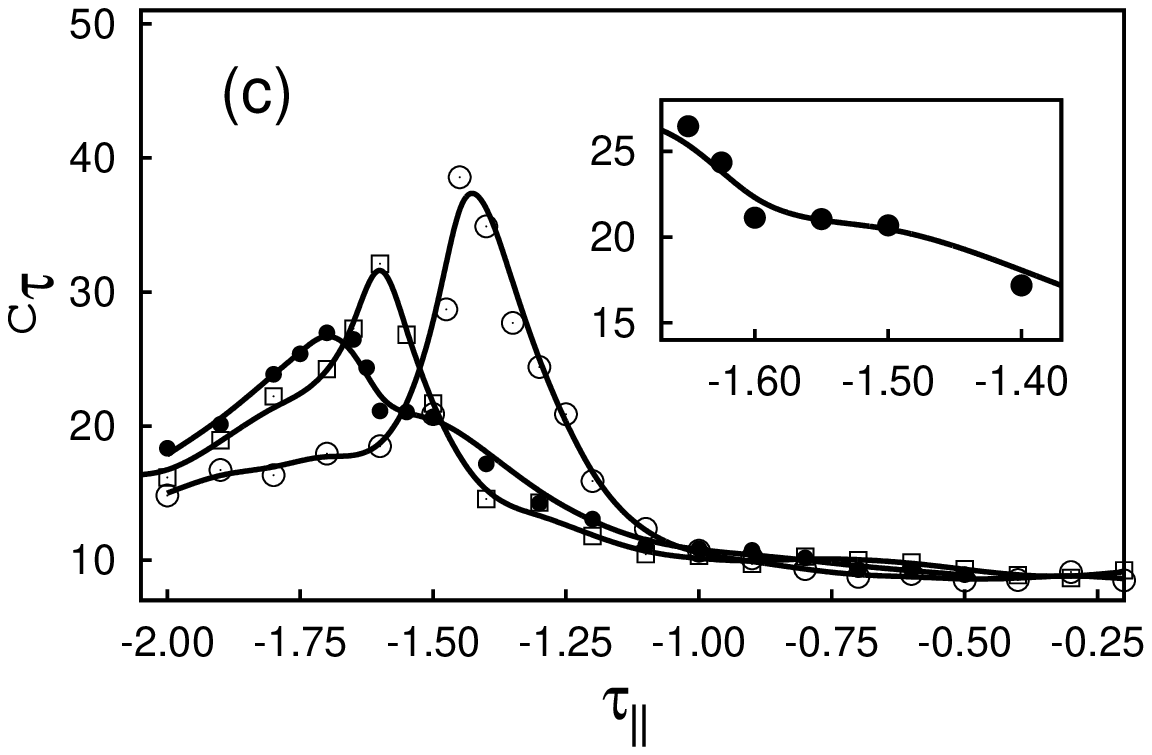,width=0.55\linewidth}
\end{center}
\caption{\small Isostress heat capacity $c_{\tau}$ as a function of compressional stress $\tau_{\parallel}$ for various homogeneous (\opensquare, \opencircle) and corresponding hybrid (\fullcircle) anchoring scenarios. (a) Homeotropic (\opensquare), planar (\opencircle). (b) Nonspecific (\opensquare), planar (\opencircle). (c) Nonspecific (\opensquare), directional (\opencircle). Insets are enlargements of plots for hybrid anchoring scenarios. Solid lines are fits intended to guide the eye.}\label{fig2}
\end{figure}

We begin our discussion with plots of $c_{\tau}$ as a function of applied compressional stress and various homogeneous and hybrid anchoring scenarios in Fig.~\ref{fig2}. The isotress heat capacity exhibits pronounced peaks at sufficiently large compressional stress. As we have already demonstrated in our previous work, maxima of $c_{\tau}$ are fingerprints of an IN transition in the confined phase. Therefore, the stresses at which these maxima are located depend on the specific homogeneous anchoring scenario. For example, the plot in Fig.~\ref{fig2}(a) shows that the IN transition occurs at lower compressional stress $\left|\tau_{\parallel}\right|\simeq1.50$ for the homogeneous homeotropic compared with $\left|\tau_{\parallel}\right|\simeq1.60$ for the homogeneous planar anchoring scenario. Moreover, these IN transition occurs at a somewhat lower stress than $\left|\tau_{\parallel}\right|\simeq1.70$ observed in the bulk under identical thermodynamic conditions (see, for example, Fig.~4.3(a) of Ref.~\citealp{greschek10}). This reflects the supportive character of the solid substrate in the formation of nematic phases. Note that in our previous study \cite{greschek10} we have been unable to observe an IN transition for the (homogeneous) homeotropic anchoring scenario because at the smaller pore width $s_{\mathrm{z}0}=10$ considered there the impact of the walls is so strong that even at vanishingly small compression almost the entire confined fluid is already in a nematic phase. As we also explained earlier the height of peaks in the plots of $c_{\tau}$ reflect a loss of rotational entropy at the IN transition which is different for different anchoring scenarios (see Sec.~V of Ref.~\citealp{greschek10}).

\begin{figure}[htb]
\begin{center}
\epsfig{file=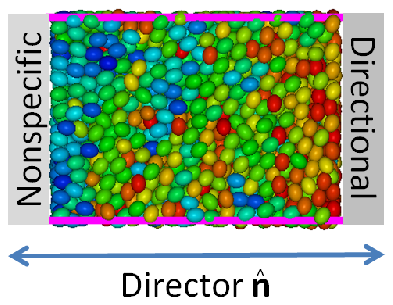,width=0.4\linewidth}
\end{center}
\caption{\small ``Snapshot'' of a representative configuration in the nematic phase. Gray boxes indicate the position of the substrates in the $x$--$z$ plane where the specific anchoring scenario is indicated within each box. The purple lines indicate the (virtual) boundaries of the simulation cell at which periodic boundary conditions are applied. To set the color code the global director is arbitrarily taken to be identical with the substrate normal. The sequence blue $\to$ green $\to$ red indicates a change in molecular orientation from parallel to perpendicular with $\widehat{\bm{n}}$ . }\label{fig3}
\end{figure}

Comparing results for the two homogeneous anchoring scenarios in Fig.~\ref{fig2}(a) with the corresponding hybrid one two main differences can be recognized. First, the dominant peak in $c_{\tau}$ is shifted to a compressional stress $\left|\tau_{\parallel}\right|\simeq1.75$ exceeding that characteristic of the two homogeneous anchoring scenarios. Obviously, the heterogeneity of the anchoring at the two substrates inhibits the formation of nematic phases. This seems quite striking at first in view of the fact that the {\em direct} impact of the solid substrates is restricted to a distance of about $\Delta z\approx4$ from either substrate surface as plots in Fig.~\ref{fig1} suggest. However, one has to keep in mind that nematic phases are characterized by {\em long-range} rotational order. Molecules of the confined soft-matter phase are therefore cooperative in transmitting the specific orientation {\em imprinted} by a solid substrate over distances substantially exceeding the range of the fluid-substrate interaction itself. The inhibition of the IN transition in the hybrid anchoring scenario may therefore be interpreted as frustration of the molecules in the confined phase which are trying to adjust their orientation with respect to the two conflicting anchoring conditions simultaneously. The structural frustration is illustrated in Fig.~\ref{fig3} where we show a ``snapshot'' of a representative molecular configuration taken from the Monte Carlo simulations.

A second interesting feature visible in the plot of $c_{\tau}$ in Fig.~\ref{fig2}(a) for the hybrid anchoring is a smaller but significant secondary peak that occurs at lower compressional stress $\left|\tau_{\parallel}\right|\simeq1.50$ compared with the main peak at $\left|\tau_{\parallel}\right|\simeq1.75$. The existence of this secondary peak has been verified in Monte Carlo runs in which we enlarged the number of cycles up to $3\times10^{6}$ cycles while carefully monitoring the evolution of $c_{\tau}$ as a function of the number of cycles similar to plots presented in Fig.~4.2 of Ref.~\citealp{greschek10}. Comparing the plots for hybrid and homogeneous homeotropic alignment in Fig.~\ref{fig2}(a) reveals that the secondary peak in the plots of $c_{\tau}$ for the hybrid anchoring coincides with that in the corresponding curve for the homogeneous homeotropic alignment. The origin of the secondary peak will become clear in Sec.~\ref{sec:locor} where we discuss local structural features of the confined liquid crystals.

\begin{figure}[htb]
\begin{center}
\epsfig{file=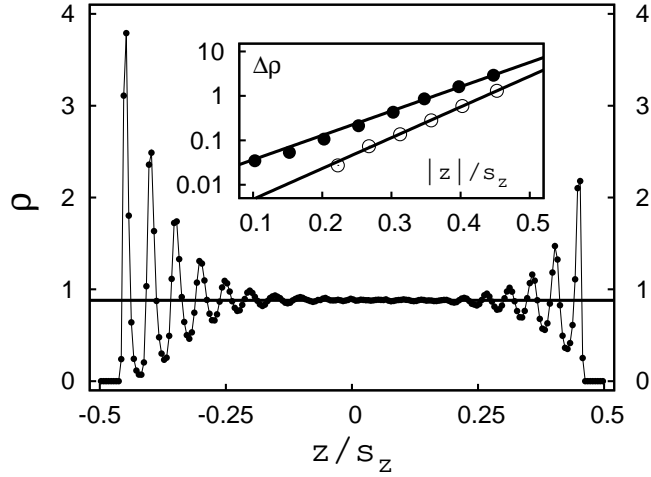,width=0.6\linewidth}
\end{center}
\caption{\small Plot of local density $\rho\left(z\right)$ as function of position between homeotropically ($z_{\mathrm{w}}/s_{\mathrm{z}}=-0.5$) and planar anchoring substrate ($z_{\mathrm{w}}/s_{\mathrm{z}}=+0.5$) at $\left|\tau_{\parallel}\right|=2.00$. The thin solid line is intended to guide the eye whereas the horizontal thick solid line corresponds to $\overline{\rho}$ in the bulk-like region of the confined phase. The inset shows a plot of $\Delta\rho$ where straight lines are fits of the far right side of eqn.~(\ref{eq:corrlen}) to $\rho\left(z_{\mathrm{m}}\right)$ (see text) near the homeotropically (\fullcircle) and planar (\opencircle) anchoring substrate.}\label{fig4}
\end{figure}

For the {\em hp} anchoring scenario we also investigated the local density $\rho\left(z\right)$ [see eqn.~(\ref{eq:rhoz})] for a state point in the nematic phase. The plot in Fig.~\ref{fig4} shows that in the confined liquid crystal molecules arrange themselves in individual layers in the vicinity of either substrate indicated by the oscillatory character of $\rho\left(z\right)$ in the vicinity of the substrates. As one moves towards the midplane of the slit-pore the fluid becomes homogeneous (i.e., bulk-like) such that $\lim\limits_{z\to0}\rho\left(z\right)\approx\overline{\rho}\simeq0.88$ is independent of $z$. As long as the confined liquid crystal is fluidic one expects
\begin{equation}\label{eq:corrlen}
\Delta\rho\equiv
\rho\left(z_{\mathrm{m}}\right)-\overline{\rho}=
a\exp\left(z/\xi_0\right)\cos\left(\xi_1z-\theta\right)
\stackrel{
{
\theta\to\pi/2
\atop
\xi_1\to0
}
}{\longrightarrow}
a\exp\left(z/\xi_0\right)
\end{equation}
where $z_{\mathrm{m}}$ is the position of maxima and $\overline{\rho}$ is the mean density of the bulk-like region in the plot of $\rho\left(z\right)$. The correlation lengths $\xi_0$ and $\xi_1$ are determined by bulk properties whereas $a$ and $\theta$ depend on the specific nature of the fluid-substrate interactions \cite{klapp08}. Note, however, that eqn.~(\ref{eq:corrlen}) holds strictly only for spherically symmetric molecules but is {\em assumed} here to be approximately valid because our system is composed of weakly anisometric molecules. A fit of eqn.~(\ref{eq:corrlen}) to $\rho\left(z_{\mathrm{m}}\right)$ gives $\xi_1\simeq0$ and $\theta\simeq\pi/2$ such that the far right side of the equation provides an excellent representation of our data (see inset in Fig.~\ref{fig4}). Moreover, the fit yields $\xi_0^{\mathrm{h}}\simeq1.5859$ near the homeotropically anchoring substrate and $\xi_0^{\mathrm{p}}\simeq1.2544$ near the planar anchoring one. The ratio $\xi_0^{\mathrm{h}}/\xi_0^{\mathrm{p}}\simeq1.26$ is nearly identical with the aspect ratio of the ellipsoidal fluid molecules as one would have anticipated.
 
As we already explained above the homogeneous nonspecific anchoring scenario favors a homeotropic alignment of the molecules in the nematic phase as well. However, as one can see from the corresponding plots in Fig.~\ref{fig2}(b) the maximum in the plot of $c_{\tau}$ for this anchoring scenario is shifted to higher compressional stress $\left|\tau_{\parallel}\right|\simeq1.60$ compared with its counterpart for homogeneous homeotropic anchoring in Fig.~\ref{fig2}(a). This indicates that the substrates for the nonspecific homogeneous anchoring scenario are somewhat less supportive in the formation of nematic phases. In fact, in the present case maxima of $c_{\tau}$ for the two homogeneous (nonspecific and planar) anchoring scenarios are located at about the same compressional stress as the plots in Fig.~\ref{fig2}(b) indicate. 

The plot of $c_{\tau}$ for the {\em np} anchoring scenario exhibits qualitatively the same features already observed in Fig.~\ref{fig2}(a) for the {\em hp} anchoring scenario. That is $c_{\tau}$ exhibits a main maximum at a larger compressional stress $\left|\tau_{\parallel}\right|\simeq1.70$ and a smaller secondary maximum at about $\left|\tau_{\parallel}\right|\simeq1.63$. The shift of the secondary peak from $\left|\tau_{\parallel}\right|\simeq1.50$ [{\em hp} anchoring, see Fig.~\ref{fig2}(a)] to $\left|\tau_{\parallel}\right|\simeq1.63$ [{\em np} anchoring, see Fig.~\ref{fig2}(b)] is more pronounced than the shift of the associated main peak in the heat-capacity curves. Notice also that the location of the secondary maximum in the plot of $c_{\tau}$ for the two hybrid anchoring scenarios in Figs.~\ref{fig2}(a) and \ref{fig2}(b) correlates nicely with the values $\left|\tau_{\parallel}\right|\simeq1.50$ and $\left|\tau_{\parallel}\right|\simeq1.60$ for the homogeneous homeotropic and nonspecific anchoring scenarios, respectively. This suggests that the secondary peak visible in the plots of $c_{\tau}$ for the two hybrid anchoring scenarios in Figs.~\ref{fig2}(a) and \ref{fig2}(b) reflects an increase in homeotropic orientation of the molecules prior to the IN transition. This notion is supported by a more detailed analysis of local structural features presented below in Sec.~\ref{sec:locor}.

\begin{figure}[htb]
\begin{center}
\epsfig{file=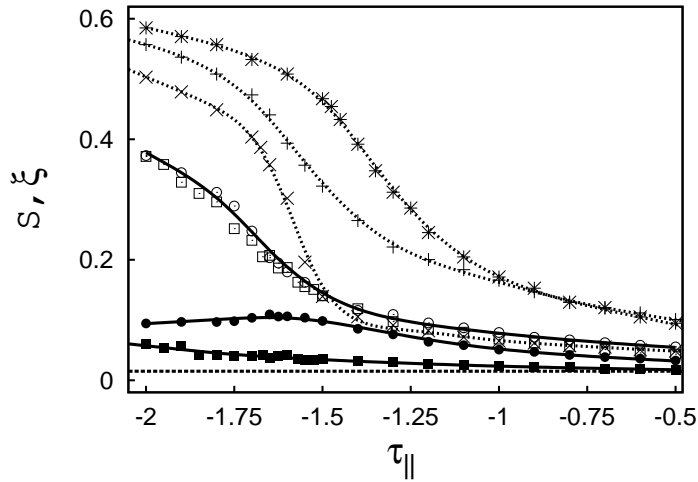,width=0.6\linewidth}
\end{center}
\caption{\small The Maier-Saupe nematic order parameter $S$ as a function of applied compressional stress. Homogeneous anchoring scenarios: ($\times$) planar, ($+$) nonspecific, ($\ast$) directional; hybrid anchoring scenarios: (\opensquare) {\em np}, (\opencircle) {\em nd}. Also shown are plots of the biaxial order parameter $\xi$ for {\em np} (\fullsquare) and {\em nd} anchoring scenarios (\fullcircle). Solid and dashed lines represent fits to guide the eye. The vertical dashed line indicates typical values of $\xi$ obtained for homogeneous anchoring scenarios.}\label{fig5}
\end{figure}

However, before turning to that discussion it is instructive to consider yet another hybrid anchoring scenario which combines nonspecific anchoring at one substrate with directional anchoring at the other one. As plots in Fig.~\ref{fig2}(c) show the maximum in $c_{\tau}$ for the {\em homogeneous} directional anchoring occurs at a compressional stress of about $\left|\tau_{\parallel}\right|\simeq1.45$ which is the smallest value of the three homogeneous anchoring scenarios considered in this study. This indicates that directional (planar) anchoring is the most supportive scenario in the formation of nematic phases in confined liquid crystals. Combining directional with nonspecific anchoring causes the small but pronounced secondary peak in the plots of $c_{\tau}$ in Figs.~\ref{fig2}(a) and \ref{fig2}(b) to give way to a broad shoulder at about $\left|\tau_{\parallel}\right|\simeq1.55$ which is located approximately halfway in between the peaks of $c_{\tau}$ at $\left|\tau_{\parallel}\right|\simeq1.45$ for the homogeneous directional and $\left|\tau_{\parallel}\right|\simeq1.60$ for the homogeneous nonspecific anchoring scenarios. One may therefore speculate that the broad shoulder in the plot of $c_{\tau}$ in Fig.~\ref{fig2}(c) reflects some other structural transformation than the secondary peaks in the plots of $c_{\tau}$ in Figs.~\ref{fig2}(a) and \ref{fig2}(b) (see Sec.~\ref{sec:locor}). 

The IN transition may also be located through plots of the (global) Maier-Saupe nematic order parameter $S$ plotted as a function of applied compressional stress in Fig.~\ref{fig5}. In all cases considered $S(\tau_{\parallel})$ is sigmoidal in shape. Taking the inflection of the sigmoidal curves as an operational definition of the location of the IN transition \cite{greschek10} very good agreement with the position of maxima in the plots of $c_{\tau}$ versus $\tau_{\parallel}$ plotted in Fig.~\ref{fig2} is obtained. For the two hybrid anchoring scenarios the curves $S(\tau_{\parallel})$ are indistinguishable which is consistent with the observation that the main peak of $c_{\tau}$ in the parallel plots in Figs.~\ref{fig2}(b) and \ref{fig2}(c) appears to be unaffected by the specific combination of anchoring conditions at the two substrates. In general, it turns out that in the nematic phase $S$ is much smaller for hybrid than for homogeneous anchoring scenarios. However, the specific hybrid anchoring scenario manifests itself in plots of the biaxial order parameter $\xi$ which turns out to be larger for the {\em nd} than for the {\em np} anchoring scenario. Also shown in Fig.~\ref{fig5} are typical values of $\xi$ for homogeneous anchoring where one expects $\xi=0$. However, in the  actual simulations one typically obtains $\xi\lesssim0.02$ on account of a small system-size effect discussed in detail in the Appendix of the paper by Eppenga and Frenkel \cite{eppenga84}.

\subsection{Local orientation}\label{sec:locor}  

\begin{figure}[htb]
\begin{center}
\epsfig{file=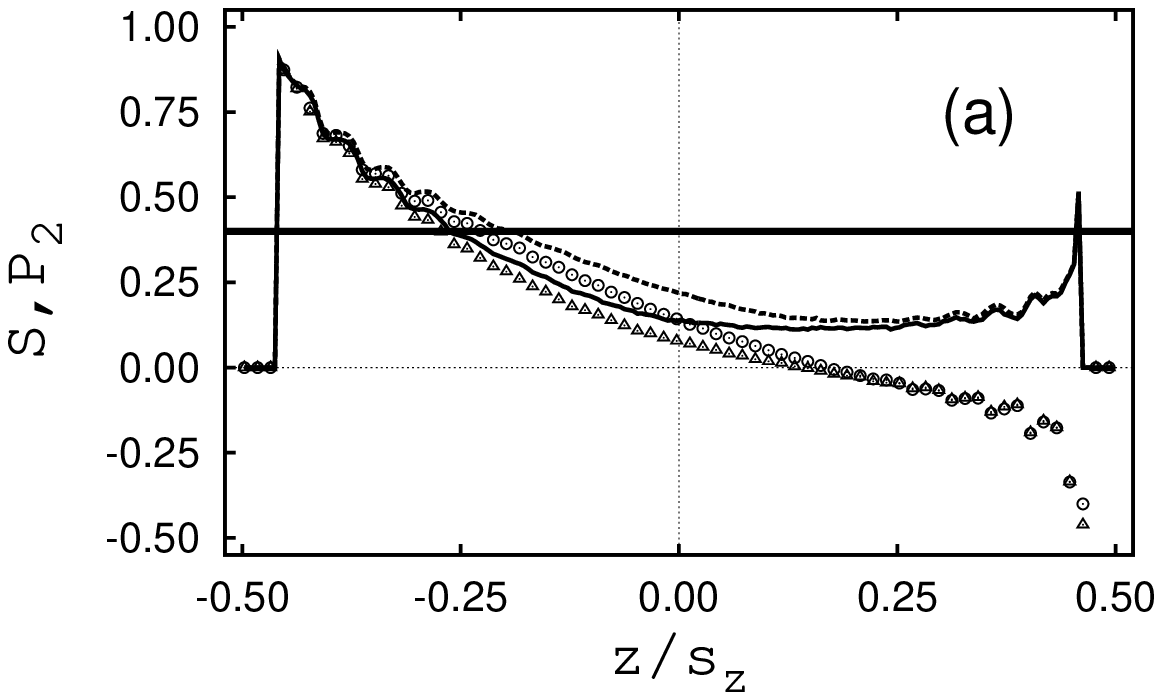,width=0.6\linewidth}
\epsfig{file=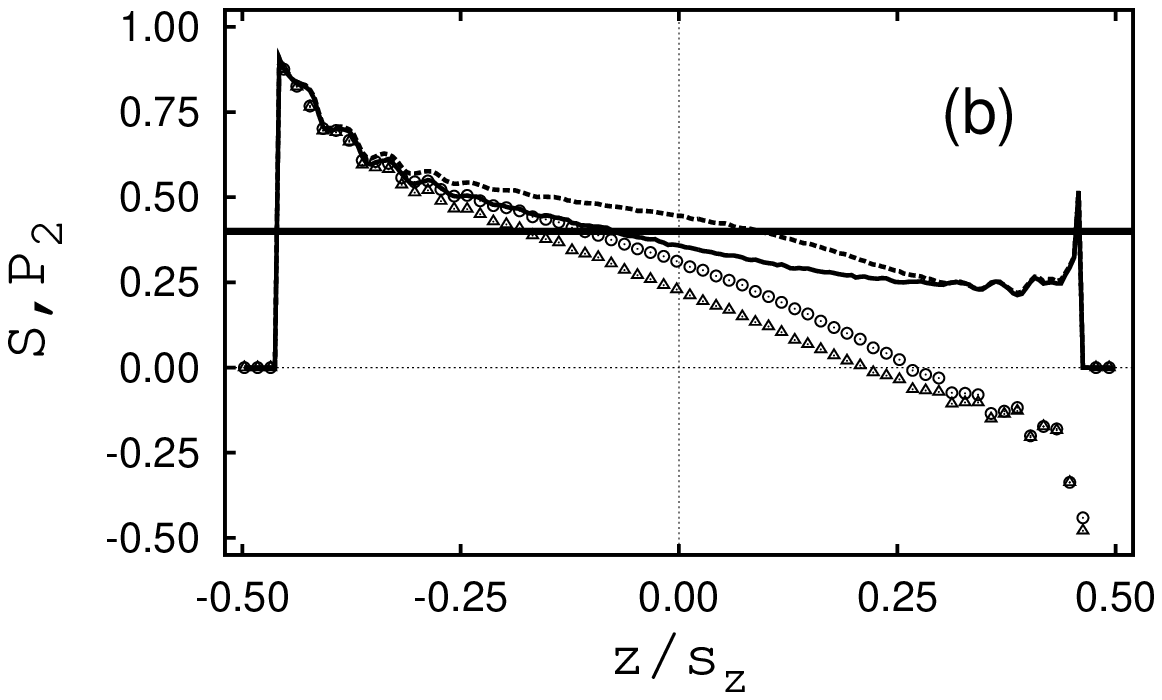,width=0.6\linewidth}
\end{center}
\caption{\small Plots of $S\left(z\right)$ (lines) and $P_2\left(z\right)$ (symbols) as functions of position between lower and upper substrates located at $z_{\mathrm{w}}/s_{\mathrm{z}0}=\mp0.5$ ($s_{\mathrm{z}0}=20$) for the {\em hp} anchoring scenario. (a) (\fullline), (\opentriangle): $\tau_{\parallel}=-1.45$; (\dashedline), (\opencircle): $\tau_{\parallel}=-1.55$. (b) (\fullline), (\opentriangle): $\tau_{\parallel}=-1.70$; (\dashedline), (\opencircle): $\tau_{\parallel}=-1.80$ [see also Fig.~\ref{fig2}(a)]. The full horizontal line demarcates $S\left(z\right)=0.4$ taken as the nematic threshold \cite{degennes95}.}\label{fig6}
\end{figure}

To gain deeper insight into structural transformations occuring in the confined liquid crystal with increasing compressional stress we consider the {\em local} Maier-Saupe nematic order parameter $S\left(z\right)$ and $P_2\left(z\right)$ [see eqn.~(\ref{eq:pz})]. According to their definition $P_2\left(z\right)=S\left(z\right)$ if molecules are aligned homeotropically with the solid substrate. If this alignment is perfect, $P_2\left(z\right)=1$. An inspection of Fig.~\ref{fig6} shows that this is the case in the immediate vicinity of the lower substrate if $g\left(\widehat{\bm{u}}\right)$ is given by eqn.~(\ref{eq:homeotropic}). If, on the other hand, the preferential orientation of molecules is in a direction parallel with the solid substrates, $P_2\left(z\right)=-0.5$ but $S\left(z\right)>0$ which is the case in the immediate vicinity of the upper substrate characterized by the anchoring function specified in eqn.~(\ref{eq:parallel}). For the plots presented in Fig.~\ref{fig6} we selected state points around the two maxima visible in the plots of $c_{\tau}$ in Fig.~\ref{fig2}(a) for hybrid anchoring. Both parts of Fig.~\ref{fig6} show that the system undergoes a structural transition during which the local order increases. These transformations are restricted mostly to ``inner'' parts of the confined fluid (i.e., regions located at $\left|z\right|/s_{\mathrm{z}0}\le0.25$) for which the direct interaction with either solid substrate is negligible according the plot in Fig.~\ref{fig1}. Notice also that in the ``nematic'' phase the local order parameter $S\left(z\right)$ is still smaller than the threshold value of $0.4$ suggested by the Maier-Saupe (mean-field) theory for the IN phase transition for $z/s_{\mathrm{z}}\gtrsim0.1$ [see Fig.~\ref{fig6}(b)]. This is the reason why the {\em global} nematic order parameter $S$ for the two hybrid anchoring scenarios plotted in Fig.~\ref{fig5} barely reaches the value $S=0.4$ even at the highest compressional stresses considered.

\begin{figure}[htb]
\begin{center}
\epsfig{file=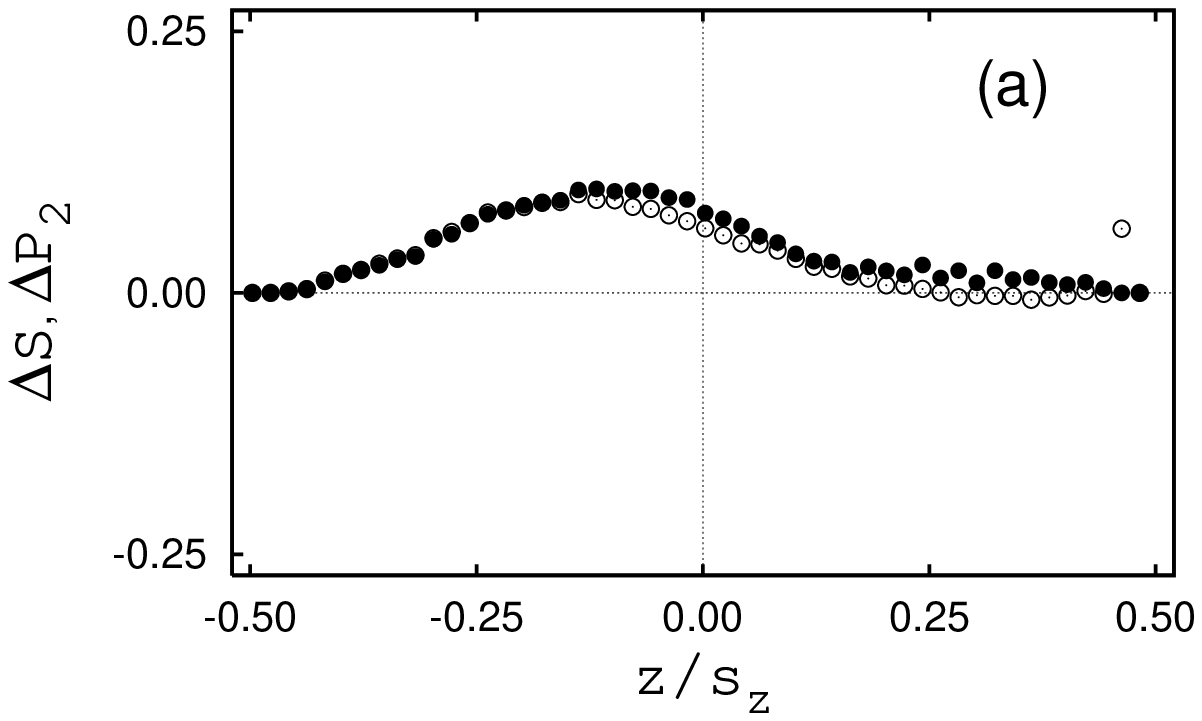,width=0.6\linewidth}
\epsfig{file=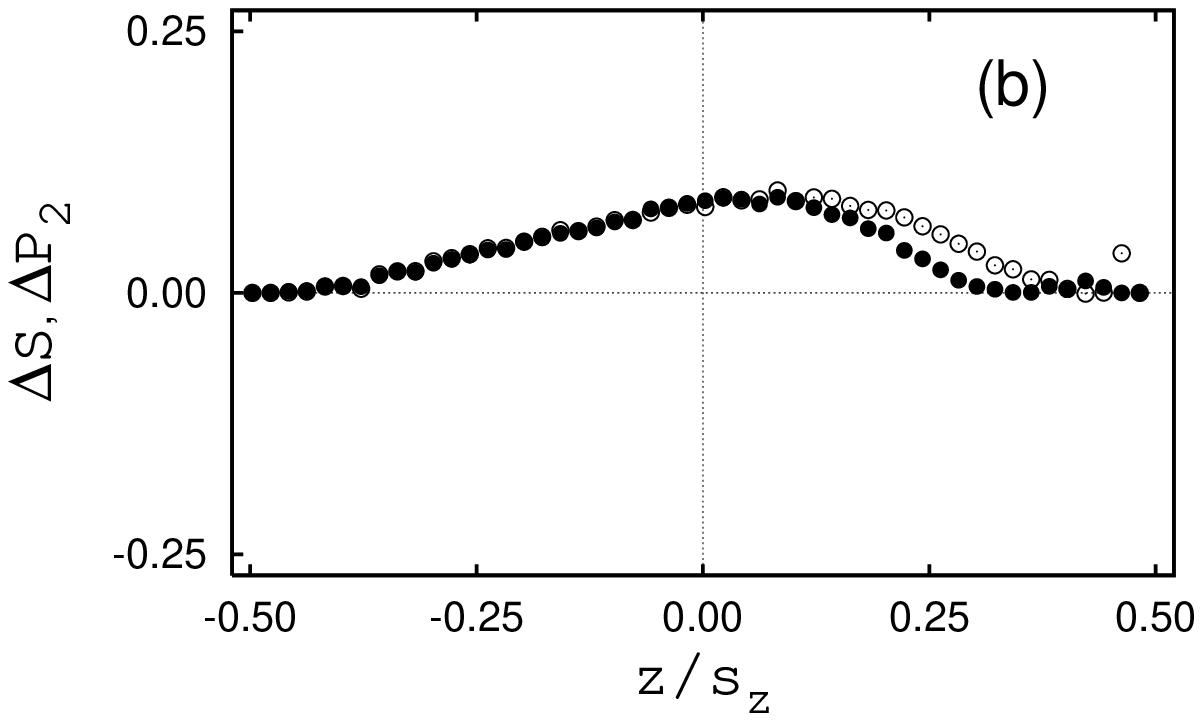,width=0.6\linewidth}
\end{center}
\caption{\small Plots of $\Delta S\left(z\right)$ (\fullcircle) and $\Delta P_2\left(z\right)$ (\opencircle) for the hybrid homeotropic-planar anchoring scenario as functions of position between lower and upper substrates. (a)  $\tau_{\parallel}^{\left(1\right)}=-1.45$, $\tau_{\parallel}^{\left(2\right)}=-1.55$, (b) $\tau_{\parallel}^{\left(1\right)}=-1.70$, $\tau_{\parallel}^{\left(2\right)}=-1.80$. Notice, that error bars may increase substantially as $\left|z\right|/s_{\mathrm{z}0}\to0.5$ because of the excluded volume in the immediate vicinity of the solid substrates (see Fig.~\ref{fig6}).}\label{fig7}
\end{figure}

To visualize subtle details of these structural transformations more clearly it turns out to be helpful to consider $\Delta S\left(z\right)\equiv S(z;\tau_{\parallel}^{\left(2\right)})-S(z;\tau_{\parallel}^{\left(1\right)})$ and, with an analogous definition, $\Delta P_2\left(z\right)$ where $|\tau_{\parallel}^{\left(1\right)}|<|\tau_{\parallel}^{\left(2\right)}|$. Plots of both quantities in Fig.~\ref{fig7}(a) show that around the secondary maximum of the heat capacity curve presented in Fig.~\ref{fig2}(a) for the hybrid anchoring scenario $\Delta S\left(z\right)=P_2\left(z\right)$ with a maximum located closer to the homeotropically anchoring substrate. For thermodynamic states around the main maximum of $c_{\tau}$ in Fig.~\ref{fig2}(a), on the other hand, plots in Fig.~\ref{fig7}(b) indicate that the maximum in the curves $\Delta S\left(z\right)$ and $\Delta P_2\left(z\right)$ is shifted towards the parallel anchoring substrate whereas $\Delta S\left(z\right)=\Delta P_2\left(z\right)$ still holds.

\begin{figure}[htb]
\begin{center}
\epsfig{file=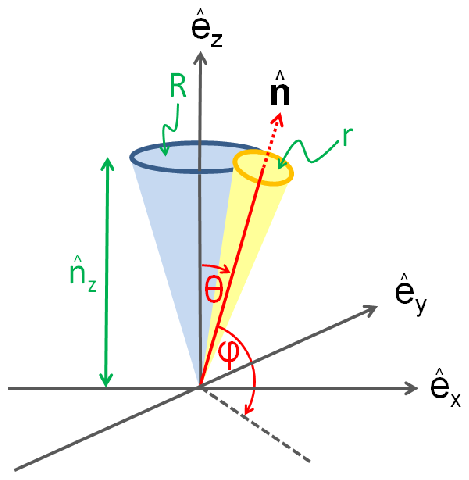,width=0.4\linewidth}
\end{center}
\caption{\small The director $\widehat{\bm{n}}$, which ends at the center of the base area of the yellow cone, in the nematic phase forms an angle $\cos\theta=\widehat{\bm{n}}\cdot\widehat{\bm{e}}_{\mathrm{z}}=\widehat{n}_{\mathrm{z}}$ with the $z$-axis, an angle $\varphi=\frac{\pi}{2}-\theta$ with the $x$--$y$ plane, and may lie anywhere on the surface of the blue cone.  Radius $R$ of the base area of the blue cone is a measure of the degree of homeotropic alignment proportional to $P_2$ [see eqn.~(\ref{eq:pz})]; radius $r$ of the base area of the yellow cone is a measure of $S$.}\label{fig8}
\end{figure}

To interpret the data presented in Fig.~\ref{fig7} it is useful to consider the schematic representation presented in Fig.~\ref{fig8}. In the nematic phase the director $\widehat{\bm{n}}$ describing the preferential orientation of the molecules will generally form a certain angle with the substrate normal (i.e., the $z$-direction) and may lie anywhere on the blue cone drawn in Fig.~\ref{fig8}. The projection $\widehat{n}_{\mathrm{z}}$ of $\widehat{\bm{n}}$ onto the $z$-axis is related to the value of $P_2$ according to the definition given in eqn.~(\ref{eq:pz}). Because $\left|\widehat{\bm{n}}\right|=1$ by definition, the radius $R$ of the circular base area of the blue cone becomes smaller and vanishes eventually if $\theta\to0$.  If, on the other hand, $\widehat{\bm{n}}$ becomes more aligned with the $x$--$y$ plane, $R$ increases and $\widehat{n}_{\mathrm{z}}$ vanishes if $\varphi\to0$. Therefore, $R$ may be viewed as a geometrical representation of the degree of homeotropic alignment. Likewise, the radius $r$ of the circular base area of the yellow cone is a geometrical representation of the Maier-Saupe nematic order parameter $S$: The larger $r$ the lower is $S$. A nonvanishing value of $r$ reflects a certain width of the distribution of molecular orientations around $\widehat{\bm{n}}$ on account of thermal fluctuations.

Applying these geometrical concepts to the plots in Fig.~\ref{fig7}(a) it is clear that $R$ and $r$ of the two cones plotted in Fig.~\ref{fig8} shrink simultaneously as the transverse compressional stress increases. This implies that the nematic order parameter $S\left(z\right)$ increases and that the director becomes more aligned with the $z$-axis. In other words, the secondary maximum in the plot of $c_{\tau}$ in Fig.~\ref{fig2}(a) at $\left|\tau_{\parallel}\right|\simeq1.50$ reflects an increasing homeotropic (pre)alignment of molecules in an isotropic region of the confined fluid [cf., Figs.~\ref{fig6}(a), \ref{fig7}(a)]. This homeotropic prealignment takes place in the part of the fluid closer to the homeotropically anchoring substrate. Similarly, the maximum at $c_{\tau}$ at $\left|\tau_{\parallel}\right|\simeq1.75$ may be interpreted as an increase in nematic order which is mostly caused by an increasing alignment of molecular orientation with the $z$-axis. However, here we note that in the vicinity of the planar anchoring substrate $\Delta S\left(z\right)<\Delta P_2\left(z\right)$. In other words, the increasing homeotropic alignment near the planar anchoring wall is associated with a slight loss of nematic order in this region. 

\begin{figure}[htb]
\begin{center}
\epsfig{file=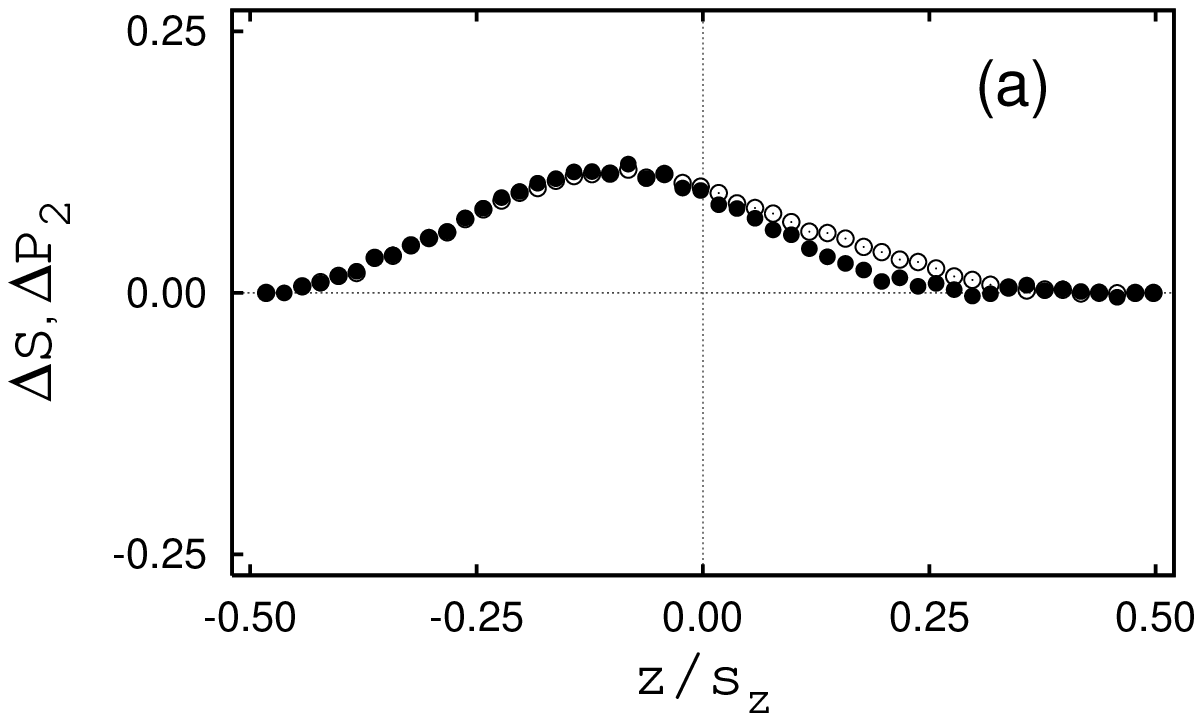,width=0.6\linewidth}
\epsfig{file=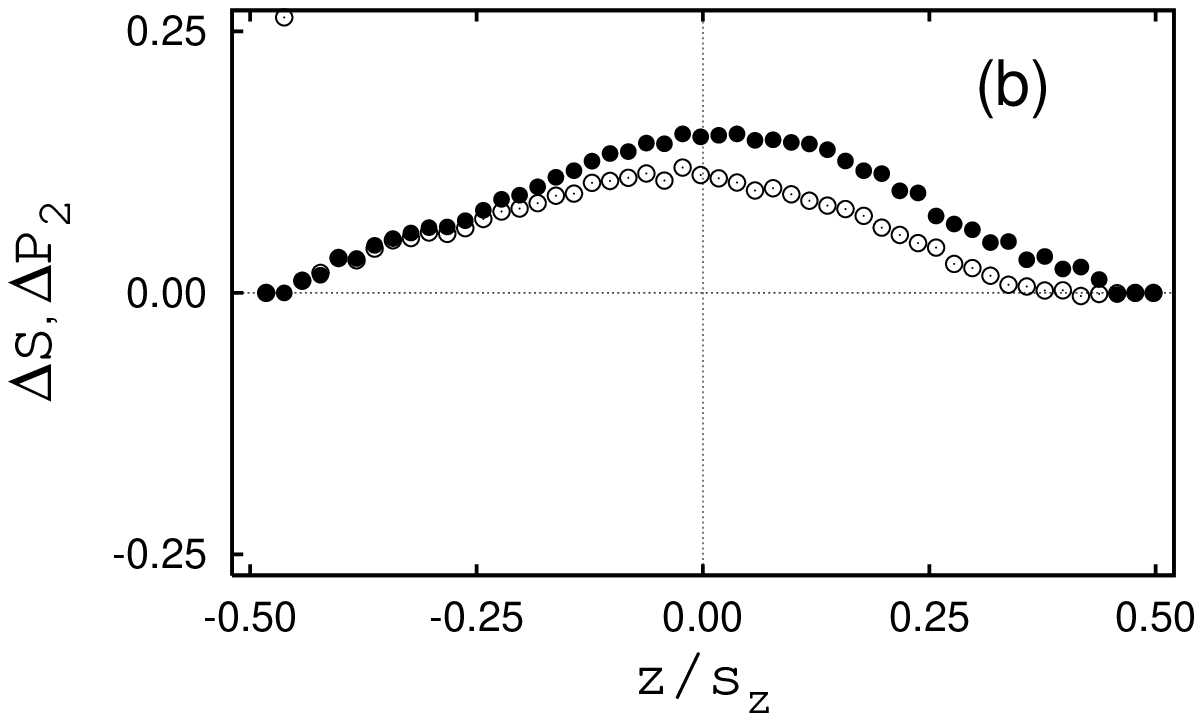,width=0.6\linewidth}
\end{center}
\caption{\small As Fig.~\ref{fig7}, but for the {\em np} anchoring scenario. (a)  $\tau_{\parallel}^{\left(1\right)}=-1.58$, $\tau_{\parallel}^{\left(2\right)}=-1.65$; (b) $\tau_{\parallel}^{\left(1\right)}=-1.67$, $\tau_{\parallel}^{\left(2\right)}=-1.80$.}\label{fig9}
\end{figure}

Corresponding plots in Fig.~\ref{fig9} for the {\em np} anchoring show that around the secondary peak of $c_{\tau}$ at $\tau_{\parallel}\simeq-1.63$ [see Fig.~\ref{fig2}(b)] the structural transformation is very similar to the one illustrated by the plots in Fig.~\ref{fig7}(a): molecules prealign preferentially with the $z$-axis in that region of the confined fluid controlled by the nonspecifically anchoring substrate. In the geometrical interpretation of Fig.~\ref{fig8} this may again be desribed in terms of a simultaneous shrinkage of both $R$ and $r$. However, the plots in Fig.~\ref{fig9}(b) illustrate a somewhat different structural transformation. Here, $\Delta S\left(z\right)>\Delta P_2\left(z\right)$ indicating that the alignment of molecules with the $z$-axis is smaller than the increase in nematic order. In the geometrical terms of Fig.~\ref{fig8} this may be desribed as a decrease of $R$ and a simultaneous larger decrease of $r$ such that $\theta$ decreases slightly in an increasingly nematic phase. 

\begin{figure}[htb]
\begin{center}
\epsfig{file=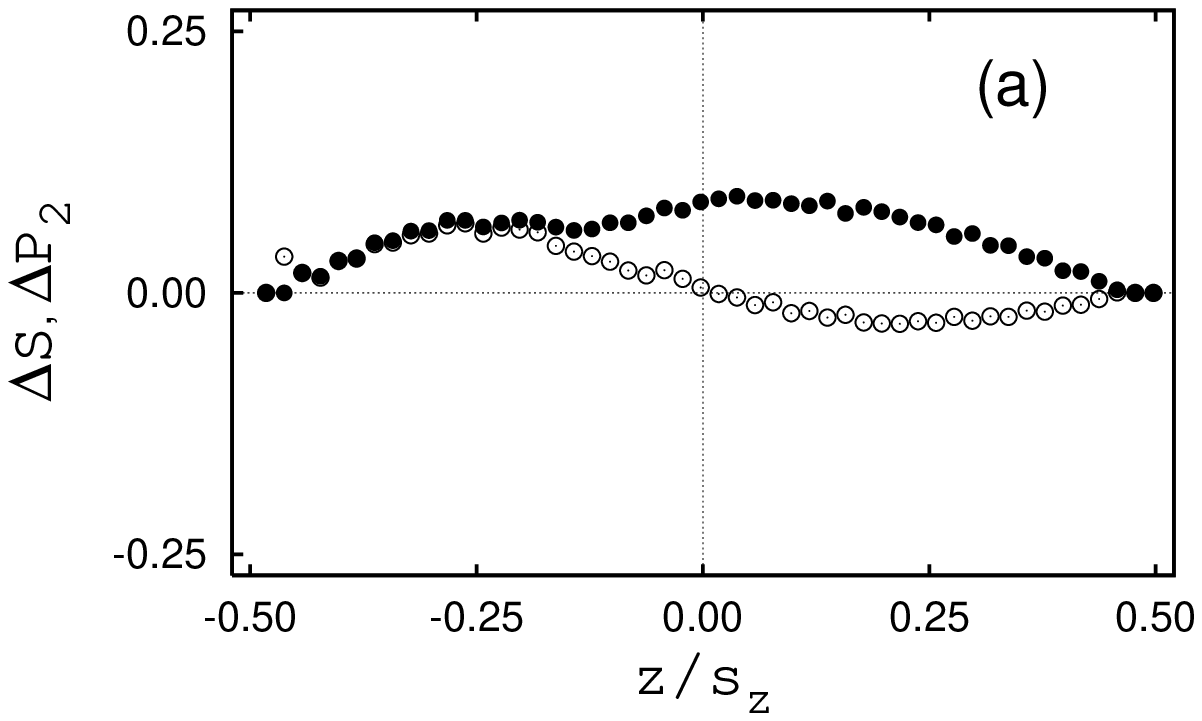,width=0.6\linewidth}
\epsfig{file=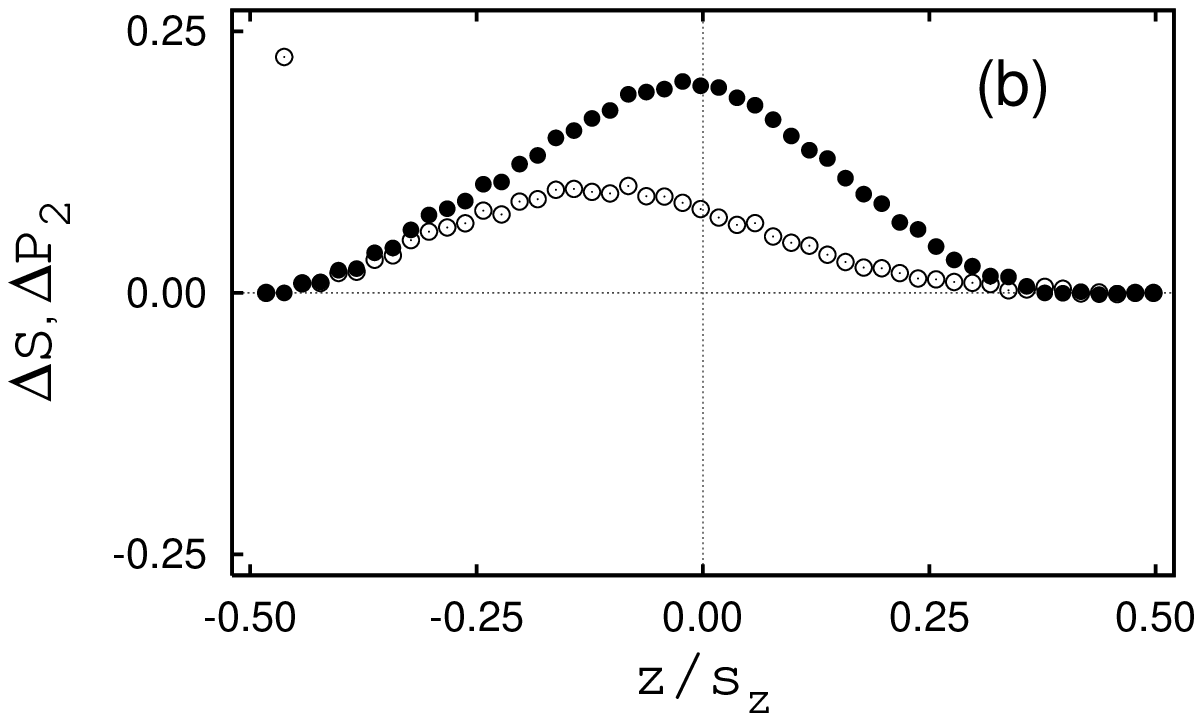,width=0.6\linewidth}
\end{center}
\caption{\small As Fig.~\ref{fig7}, but for the {\em nd} anchoring scenario. (a)  $\tau_{\parallel}^{\left(1\right)}=-1.50$, $\tau_{\parallel}^{\left(2\right)}=-1.60$; (b) $\tau_{\parallel}^{\left(1\right)}=-1.63$, $\tau_{\parallel}^{\left(2\right)}=-1.75$.}\label{fig10}
\end{figure}

A totally different structural transformation is observed for the {\em nd} anchoring scenario for which we present plots of $\Delta S\left(z\right)$ and $\Delta P_2\left(z\right)$ in Fig.~\ref{fig10}. For two thermodynamic states in the vicinity of the weak shoulder in the plot of $c_{\tau}$ at $\left|\tau_{\parallel}\right|\simeq1.55$ [see Fig.~\ref{fig2}(c)] the plots in Fig.~\ref{fig10}(a) indicate that in the immediate vicinity of the hybrid anchoring substrate $\Delta S\left(z\right)=\Delta P_2\left(z\right)$ referring to an enahnced homeotropic alignment of the molecules as the compressional stress increases. On the contrary, in the upper halfspace of the slit-pore (i.e., for $z/s_{\mathrm{z}0}\ge0$),  $\Delta P_2\left(z\right)<0$ whereas $\Delta S\left(z\right)>0$. In the geometric terms introduced in Fig.~\ref{fig8} this can be ascribed to an increase of $R$ (i.e., a decrease of $\widehat{n}_{\mathrm{z}}$) and a decrease of $r$. In other words, the increase in nematic order is a consequence of a preferential alignment of the molecules with the $x$-axis. At higher compressional stresses in the vicinity of the main peak of $c_{\tau}$ at $\left|\tau_{\parallel}\right|\simeq1.70$ [see Fig.~\ref{fig2}(c)] this trend is somewhat reversed as the corresponding plots in Fig.~\ref{fig10}(b) show. The plot of $\Delta S\left(z\right)$ is symmetric about the midplane of the slit-pore and because $\Delta S\left(z\right)>0$ the nematic order in the system increases mainly at the center of the confined fluid. There is also some enhanced homeotropic alignment which turns out to be somewhat more pronounced in the vicinity of the homeotropically anchoring wall in accord with one's physical intuition.
\setcounter{figure}{0}
\section{Summary and conclusions}\label{sec:sumcon}
In this work we consider a simple model of a liquid crystal confined to a slit-pore with hybrid anchoring at the solid substrates, that is each substrate discriminates molecules on account of their orientation with respect to the substrate plane where we take this energetic discrimination to be different at the two substrates. Employing Monte Carlo simulations in a specialized isostress ensemble we can locate specific compressional stresses, at which the fluid undergoes a significant structural reorganization, through characteristic maxima of the isostress heat capacity. 

As in our previous study \cite{greschek10} the location of these maxima coincides with inflection points in plots of the Maier-Saupe nematic $S$ order parameter as a function of compressional stress. Considering various hybrid anchoring scenarios it turns out that the location of the IN transition is largely independent of the combination of anchoring conditions at the two substrates. This is because in the nematic phase the director is trying to point in a direction compromising between the competing orientations enforced by the two substrates simultaneously. Consequently, plots of $S$ versus $\tau_{\parallel}$ are nearly indistinguishable as far as the {\em np} and {\em nd} anchoring scenarios are concerned. Our parallel analysis of local order parameter profiles shows that the IN transformation occurs near the center of the confined liquid crystal that is approximately halfway in between the solid substrates.

However, the specific hybrid anchoring scenario does have an impact on a secondary structural reorganization that occurs at lower compressional stresses preceding that of the IN transition. This secondary transformation may cause prealignment with the $z$-axis (i.e., in a direction perpendicular to the substrate plane) or it may force molecules to prealign to a certain extent with the $x$--$y$ plane. Which one of these two cases is realized depends on the stronger anchoring scenario in the hybrid pair. Thus, our results show that homeotropic is stronger than planar anchoring during this prealignment process. The same is true for the nonspecific anchoring which favors a homeotropic alignment of the fluid molecules on account of the symmetry breaking presence of the solid substrates. In the combination with nonspecific anchoring the directionally anchoring substrate ``wins'' and is thus considered stronger in the above sense.

In slit-pores with hybrid anchoring at the substrates one generally anticipates a certain degree of biaxiality. Our simulations do, however, show that the degree of biaxiality is actually quite small. Only the {\em nd} anchoring scenario exhibits a significant degree of biaxiality of about $25\%$ of $S$ in the nematic phase. The apparent lack of a more substantial biaxiality can be understood by realizing that the director field $\widehat{\bm{n}}\left(z\right)$ varies as a function of position w.r.t. the location of the substrate plane. This is inferred from a characteristic ``snapshot'' of a configuration of molecules in the nematic phase of the {\em nd} anchoring scenario. In other words, the preferential orientation of molecules in the nematic phase varies with position such that more than just two preferred orientations in the confined liquid crystal are present. The small nonvanishing biaxiality for the {\em nd} anchoring scenario is caused by those few layers of molecules in the vicinity of each substrate that exhibit either a homeotropic or a planar orientation of the molecules.     

\appendix
\setcounter{figure}{0}
\section{Perturbational treatment of biaxial symmetry}\label{sec:appa}
In this appendix we derive eqn.~(\ref{eq:biax}) by perturbational arguments. We assume that we can split the alignment tensor $\mathbf{Q}$ for a system with biaxial symmetry into a contribution $\mathbf{Q}^{\left(0\right)}$ representing a system with uniaxial symmetry (i.e., the ``symmetry ground state'') and a perturbation $\mathbf{Q}^{\left(1\right)}$ which removes the rotational symmetry in the plane orthogonal to the symmetry axis in the uniaxial case (``symmetry excited state''). Hence, we may write
\begin{equation}\label{eq:split}
\mathbf{Q}=\mathbf{Q}^{\left(0\right)}+\mathbf{Q}^{\left(1\right)}
\end{equation}
where $\mathbf{Q}$, $\mathbf{Q}^{\left(0\right)}$, and $\mathbf{Q}^{\left(1\right)}$ are assumed to be represented by symmetric, traceless, $3\times3$ matrices. Let $\mathbf{Q}^{\left(0\right)}$ satisfy the eigenvalue equation
\begin{equation}\label{eq:eigen}
\mathbf{Q}^{\left(0\right)}\widehat{\bm{x}}_i^{\left(0\right)}=\lambda_i^{\left(0\right)}\widehat{\bm{x}}_i^{\left(0\right)},\qquad i=1,\ldots,3
\end{equation}
where $\{\widehat{\bm{x}}_i^{\left(0\right)}\}$ and $\{\lambda_i^{\left(0\right)}\}$ are the sets of eigenvectors and -values of $\mathbf{Q}^{\left(0\right)}$, respectively where we introduce matrix elements $Q_{ij}^{\left(0\right)}$ of $\mathbf{Q}^{\left(0\right)}$ via $Q_{ij}^{\left(0\right)}\equiv\widehat{\bm{x}}_i^{\left(0\right)}\mathbf{Q}^{\left(0\right)}\widehat{\bm{x}}_j^{\left(0\right)}$. Because $\mathbf{Q}^{\left(0\right)}$ is real and symmetric, $Q_{ij}^{\left(0\right)}=Q_{ji}^{\left(0\right)}$. It then follows directly from eqn.~(\ref{eq:eigen}) that
\begin{equation}\label{eq:ortho}
0=
\left(
\lambda_i^{\left(0\right)}-\lambda_j^{\left(0\right)}
\right)
\widehat{\bm{x}}_i^{\left(0\right)}\cdot\widehat{\bm{x}}_j^{\left(0\right)}
\end{equation}
which expresses the well-known fact that eigenvectors to different eigenvalues are automatically orthogonal to one another. Let $\lambda_3^{\left(0\right)}=\lambda_{\mathrm{m}}$ be the largest eigenvalue of $\mathbf{Q}^{\left(0\right)}$. Uniaxial symmetry then requires $\lambda_1^{\left(0\right)}=\lambda_2^{\left(0\right)}=\lambda^{\prime}\le\lambda_{\mathrm{m}}$ implying $\widehat{\bm{x}}_i^{\left(0\right)}\cdot\widehat{\bm{x}}_3^{\left(0\right)}=0$ ($i=1,2$), that is eigenvector $\widehat{\bm{x}}_3^{\left(0\right)}$ is orthogonal to the other two. However, because $\lambda_1^{\left(0\right)}=\lambda_2^{\left(0\right)}$, $\widehat{\bm{x}}_1^{\left(0\right)}$ and $\widehat{\bm{x}}_2^{\left(0\right)}$ can lie anywhere in the plane orthogonal to $\widehat{\bm{x}}_3^{\left(0\right)}$. In particular, eqn.~(\ref{eq:ortho}) remains valid without requiring these two eigenvectors to be orthogonal to one another. Hence, only $\widehat{\bm{x}}_3^{\left(0\right)}$ defines a distinct direction and represents the axis of uniaxial symmetry.

To proceed it is important to realize that as long as the original eigenvectors are linearly independent one may apply a standard orthogonalization procedure (such as, for example, Gram-Schmidt, see Ref.~\citealp{arfken85}) to replace $\{\widehat{\bm{x}}_i^{\left(0\right)}\}$ by $\{\widehat{\bm{y}}_i^{\left(0\right)}\}$ where now $\widehat{\bm{y}}_i^{\left(0\right)}\cdot\widehat{\bm{y}}_j^{\left(0\right)}=\delta_{ij}$, $\forall i,j$. Moreover, members of the new set $\{\widehat{\bm{y}}_i^{\left(0\right)}\}$ are still eigenvectors of $\mathbf{Q}^{\left(0\right)}$ to the same eigenvalues $\lambda^{\prime}$ and $\lambda_{\mathrm{m}}$. Because $\{\widehat{\bm{y}}_i^{\left(0\right)}\}$ form a complete orthonormal set of eigenvectors one may construct a matrix $\mathbf{Y}_0$ where column $i$ is formed by $\widehat{\bm{y}}_i^{\left(0\right)}$. With this matrix $\mathbf{Q}^{\left(0\right)}$ may be diagonalized according to
\begin{equation}\label{eq:diagQ0}
\diag\mathbf{Q}^{\left(0\right)}=
\bm{\Lambda}_0=
\mathbf{Y}_0^{-1}\mathbf{Q}^{\left(0\right)}\mathbf{Y}_0=
\left(
\begin{array}{ccc}
\lambda^{\prime}&0&0\\
0&\lambda^{\prime}&0\\
0&0&\lambda_{\mathrm{m}}
\end{array}
\right)=
\left(
\begin{array}{ccc}
-\lambda_{\mathrm{m}}/2&0&0\\
0&-\lambda_{\mathrm{m}}/2&0\\
0&0&\lambda_{\mathrm{m}}
\end{array}
\right)
\end{equation} 
where the far right side follows because the trace is an invariant of a matrix such that $\Tr\mathbf{Q}^{\left(0\right)}=\Tr\bm{\Lambda}_0=0$.

Assume now that $\mathbf{Q}$ satisfies an eigenvalue equation similar to eqn.~(\ref{eq:eigen}) that is 
\begin{equation}\label{eq:eigenQ}
\mathbf{Q}\widehat{\bm{y}}=\lambda\widehat{\bm{y}}
\end{equation}
where we may express the eigenvector $\widehat{\bm{y}}$ in the basis $\{\widehat{\bm{y}}_i^{\left(0\right)}\}$ according to
\begin{equation}\label{eq:linsup}
\widehat{\bm{y}}=
a_1\widehat{\bm{y}}_1^{\left(0\right)}+
a_2\widehat{\bm{y}}_2^{\left(0\right)}+
a_3\widehat{\bm{y}}_3^{\left(0\right)}
\end{equation}
We determine the expansion coefficients $\{a_i\}$ by inserting eqn.~(\ref{eq:linsup}) into eqn.~(\ref{eq:eigenQ}) and multiply the resulting expression from the left successively by $\widehat{\bm{y}}_i^{\left(0\right)}$, $i=1,\ldots,3$. Employing the orthonormality of the basis functions we obtain a set of three coupled linear equations which has a nontrivial solution if and only if the secular determinant
\begin{equation}\label{eq:secdet}
\left|
\begin{array}{ccc}
-\lambda_{\mathrm{m}}/2-\lambda&\pm\zeta&0\\
\pm\zeta&-\lambda_{\mathrm{m}}/2-\lambda&0\\
0&0&\lambda_{\mathrm{m}}-\lambda
\end{array}
\right|\stackrel{!}{=}0
\end{equation}
The specific form of the secular determinant follows from eqn.~(\ref{eq:split}) and (\ref{eq:eigen}) assuming that
\begin{subequations}\label{eq:perturb}
\begin{eqnarray}
\mathbf{Q}^{\left(1\right)}
\widehat{\bm{y}}_i^{\left(0\right)}&=&
\pm\zeta\widehat{\bm{y}}_j^{\left(0\right)},\qquad i\ne j=1,2\label{eq:perturb12}\\
\mathbf{Q}^{\left(1\right)}
\widehat{\bm{y}}_3^{\left(0\right)}&=&0\label{eq:perturb3}
\end{eqnarray}
\end{subequations}
where $\zeta$ is the biaxiality parameter. We have chosen the perturbation such that it preserves the orthonormality of the basis function but removes the degeneracy of the eigenstates in the plane orthogonal to $\widehat{\bm{x}}_3^{\left(0\right)}=\widehat{\bm{y}}_3^{\left(0\right)}$ by introducing a second length scale represented by $\zeta$. This can be seen more directly by computing the roots of the third-order polynomial in $\lambda$ represented by eqn.~(\ref{eq:secdet}) which turn out to be given by
\begin{subequations}\label{eq:eigenvalQ}
\begin{eqnarray}
\lambda_0&=&\lambda_{\mathrm{m}}\\
\lambda_{\pm}&=&-\frac{\lambda_{\mathrm{m}}}{2}\pm\zeta
\end{eqnarray}
\end{subequations}
where the associated eigenvectors of $\mathbf{Q}$ are given by
\begin{subequations}\label{eq:eigenvecQ}
\begin{eqnarray}
\widehat{\bm{y}}_0&=&\widehat{\bm{y}}_3^{\left(0\right)}\\
\widehat{\bm{y}}_{\pm}&=&\frac{1}{\sqrt{2}}
\left(
\widehat{\bm{y}}_1^{\left(0\right)}\pm\widehat{\bm{y}}_2^{\left(0\right)}
\right)
\end{eqnarray}
\end{subequations}
Using the eigenvectors $\widehat{\bm{y}}_0$ and $\widehat{\bm{y}}_{\pm}$ as a new basis we may set up a matrix $\mathbf{Y}$ analogous to $\mathbf{Y}_0$ in eqn.~(\ref{eq:diagQ0}) and diagonalize $\mathbf{Q}$ according to
\begin{equation}\label{eq:diagQ}
\diag\mathbf{Q}=
\mathbf{Y}^{-1}\mathbf{Q}\mathbf{Y}=
\left(
\begin{array}{ccc}
-\lambda_{\mathrm{m}}/2+\zeta&0&0\\
0&-\lambda_{\mathrm{m}}/2-\zeta&0\\
0&0&\lambda_{\mathrm{m}}
\end{array}
\right)
\end{equation}

\acknowledgments
We are grateful for financial support from the International Graduate Research Training Group 1524 ``Self-assembled soft-matter nanostructures at interfaces''. 

\bibliographystyle{rsc-bibtex}
\bibliography{manuscript}

\end{document}